\documentclass[11pt,reprint,showpacs,showkeys,amsfonts,amssymb,amsmath,nofootinbib]{article}
\usepackage{bm,amsmath,epsf,epsfig,latexsym,amssymb,cite,lmodern}
\usepackage{graphics,psfrag,hyperref}
\usepackage[utf8]{inputenc}

\newcommand{\Ima}{{\rm Im}\,}

\begin{document}

\thispagestyle{empty}


\title{ Testing the $\chi_{c1}\, p$ composite nature of the $P_c(4450)$}


\author{Ulf-G. Mei\ss ner$^{1,2,}$\footnote{Email address:
        \texttt{meissner@hiskp.uni-bonn.de} },~
        Jos\'e~A.~Oller$^{3,}$\footnote{Email address:
        \texttt{oller@um.es} }\\[0.5em]
        {\it\small $^1$ Helmholtz-Institut f\"ur Strahlen- und Kernphysik and 
                        Bethe Center for Theoretical Physics,}
        \\{\it\small Universit\"at Bonn, D-53115 Bonn, Germany}\\
        {\it\small $^2$Institute for Advanced Simulation, Institut f\"ur Kernphysik
          and J\"ulich Center for Hadron Physics,}\\
        {\it\small  Forschungszentrum J\"ulich, D-52425 J\"ulich, Germany}\\
        {\it\small $^3$Departamento de F\'{\i}sica. Universidad de Murcia. E-30071 Murcia, Spain}}


\maketitle

\begin{abstract}
Making use of a recently proposed formalism, we analyze the composite nature
of the $P_c(4450)$ resonance observed by LHCb. We show
that the present data  suggest that this state is almost entirely made of a 
$\chi_{c1}$ and a proton, due to the close proximity to this threshold. This
also suppresses the decay modes into other, lighter channels, in our study
represented by $J/\Psi p$. We further argue that this is very similar to the 
case of the scalar meson $f_0(980)$ which is located closely to the $K\bar K$
threshold and has a suppressed decay into the lighter $\pi\pi$ channel.
\end{abstract}


\newpage

\section{Introduction}
\label{sec:250715.1}

A clear peak has been observed in the  invariant mass distribution of the $J/\Psi p$ subsystem in 
the three-body decay $\Lambda_b\to K^- J/\Psi p$ in Ref.~\cite{ref.230715.1}. 
 This signal is interpreted as a new pentaquark resonance
 $P_c(4450)$. These results have spurned a plethora of theoretical investigations, see
 Refs.~\cite{Chen:2015loa,Chen:2015moa,Roca:2015dva,Feijoo:2015cca,Mironov:2015ica,Guo:2015umn,Maiani:2015vwa,He:2015cea,Liu:2015fea,Lebed:2015tna}.
Clearly, it is still necessary to obtain further experimental information to confirm 
that it is actually a resonance and not an anomalous-threshold singularity  on top of the 
$\chi_{c1} p$ branch-point threshold \cite{Guo:2015umn}. Other proposals like in Ref.~\cite{Roca:2015dva} 
consider the $P_c(4450)$ as a composite of much heavier channels,  with some
of them having a large width. Here, we further explore the ideas
first given in Ref.~\cite{Guo:2015umn}, as the extreme close proximity of the
mass of the $P_c(4450)$ to the $\chi_{c1} p$ threshold can not be accidental.
Independent of its true nature, to simplify the argumentation, we
employ the term resonance to indicate the peak $P_c(4450)$ unveiled  in Ref.~\cite{ref.230715.1}. 

We proceed by analogy with the $J^{PC}=0^{++}$ resonance $f_0(980)$ which couples mainly 
to two channels, one lighter ($\pi\pi$) and another heavier ($K\bar{K}$), with
the latter threshold almost coinciding with the  resonance mass. The lighter channel
is the one that drives  the relatively small width of the $f_0(980)$, in the
sense that it is responsible for this resonance to develop a width,
 although it owes its origin  to the nearby $K\bar{K}$ threshold \cite{ref.240715.1}. 
In our present case for the $P_c(4450)$ resonance, the $J/\Psi p$ state is
assumed to play the role  of the $\pi\pi$ channel in the case of the
$f_0(980)$ because, despite the resonance having plenty of phase space to decay 
into this channel, the $P_c(4450)$ width is rather small. This indicates that
the coupling to this channel  is suppressed. Following this line of reasoning,
the  $\chi_{c1} p$ channel  is assumed to play the analogous role of the
$K\bar{K}$ one for the $f_0(980)$,  because it is almost on 
top of the mass of the $P_c(4450)$,  dragging  the mass of the resonance
towards its threshold due to its large coupling, being furthermore the main 
contribution in the resonance composition.

In the following, we work out the consequences of such a scenario making use
of the formalism developed in Ref.~\cite{ref.230715.3}. This allows to 
calculate the compositeness coefficients for the involved channels and also
the partial decay widths. As we will show, the present data can be well
described with the assumption that the $P_c(4450)$ is a $\chi_{c1} p$
composite, either a resonance or an anomalous threshold singularity. How to
exclude the latter scenario was already discussed in Ref.~\cite{Guo:2015umn}.
In particular, a precise measurement of the partial widths would be a clear test
of this scenario of the composite nature of the  $P_c(4450)$.

\section{Compositeness condition and the width of the $P_c(4450)$}
\label{sec:230715.1}

As stated above, we consider a two-channel scenario, where the extreme
proximity of the $P_c (4450)$ to the second channel, $\chi_{c1} p$, suggests
that the resonance is in fact a composite of this latter constituent state, 
suppressing in this way the otherwise large width to the lighter mass
channel $J/\Psi p$.  
 To further investigate this possibility, we make use of the recent work
from Ref.~\cite{ref.230715.3} that established a well defined procedure to 
interpret in a standard probabilistic way the compositeness of a resonance 
from the compositeness relation  
\cite{Weinberg:1962hj,Baru:2003qq,ref.230715.4,ref.230715.5,ref.230715.6,Agadjanov:2014ana}. 
The final result is a  simple prescription consisting  in changing the phase
of every coupling separately, such that the compositeness coefficient of the 
corresponding  channel is a positive real number. In this way, if the original 
couplings are denoted by $\gamma^i$,  the procedure of
Ref.~\cite{ref.230715.3} requires to transform them as
\begin{align}
\label{230715.1}
\gamma_i\to \gamma_i e^{i\varphi_i}~.
\end{align}
This change results from the  determination of  the physically suited 
unitary transformation  of the $S$-matrix.    
 Each of these unitary transformations 
implies a new compositeness relation, all of them being associated with the
same resonance projection operator ${\cal A}$.  A detailed account of this
theory is developed in Ref.~\cite{ref.230715.3}.  In this 
way the compositeness coefficient $X_i$ transform as
\begin{align}
\label{230715.2}
X_i=&-\gamma_i^2 \frac{\partial G_i(s_P)}{\partial s}
\to 
X_i^f= -e^{2i\varphi_i} \gamma_i^2 \frac{\partial G_i(s_P)}{\partial s} =|X_i|~,
\end{align}
where $X_i^f$ is the final compositeness coefficient for channel $i$,
 representing the amount of this  two-body  channel in the resonance state.
 In the previous equation, $s_P$ is the pole position of the resonance and 
 $G_i(s)$ is the scalar unitary loop function for channel $i$, which requires 
a subtraction constant,  see e.g. Refs.~\cite{Oller:2000fj,ref.230715.7} 
for explicit expressions of this function. 
But since only the derivative of $G_i$ enters in $|X_i|$, this coefficient 
is independent of the subtraction constant. The derivative of $G_i(s)$ is 
a definite function of $s$ and the precise 
values of the masses of the particles involved, as  
it would correspond to a convergent three-point one-loop function.

Our basic criterion is to impose that the $P_c(4450)$ is a composite mainly 
of $\chi_{c1}p$ with some  contribution from the $J/\Psi p$. Indeed, if we
assumed that this last channel were the main 
contribution in  the composition of the resonance (in the following we
indicate by 1 the channel $J/\Psi p$  and by 2 the $\chi_{c1}p$, in order of
increasing thresholds) then from the requirement
\begin{align}
\label{230715.3}
|X_1|\simeq 1 
\end{align}
it would follow that
\begin{align}
\label{230715.4}
|\gamma_1|^2=\left|\frac{\partial G_1(s_P)}{\partial s}\right|^{-1}~.
\end{align}
 We determine  $s_P$ from the values 
of the mass, $m_P$, and  width, $\Gamma_P$, of the $P_c(4450)$  obtained in Ref.~\cite{ref.230715.1}, as
\begin{align}
\label{240715.1}
s_P=(m_P-\frac{i}{2}\Gamma_R)^2~.
\end{align}
Performing this simple exercise one would then obtain a huge and completely 
unrealistic width 
\begin{align}
\label{240715.2}
\Gamma_P=&\frac{q_1(m_P^2) |\gamma_1|^2}{8\pi m_P^2}\simeq 1.5~\text{GeV},
\end{align}
where  we denote by  $q_i(s)$ the three-momentum of channel $i$ at $s$ (the
total energy squared in the center-of-mass frame). 
Note that the threshold of  $J/\Psi p$ is around 400~MeV lighter than the
nominal mass of the resonance $P_c(4450)$ so that 
there is plenty of phase space to allow the previous decay.\footnote{Because
  of this, one should take $s_P$ in the unphysical  Riemann sheet for this channel 
when evaluating the derivative of $G_1$ at $s_P$.}   
This is similar to the case of the $f_0(980)$ resonance where the $\pi\pi$
channel is required to couple  weakly to the $f_0(980)$, otherwise the width
of the latter would be  huge.

Let us now proceed to reach quantitative conclusions within our working
assumption, consisting in the analogy,  on the one hand,  between the channels
$\pi\pi$, $K\bar{K}$  and  $J/\Psi p$, $\chi_{c1} p$,  in order, and on the
other hand, between the resonances $f_0(980)$ and $P_c(4450)$. Our main equations 
stem from  imposing  saturation of the compositeness relation  and the width
of the  $P_c(4450)$  by the channels  considered,  $J/\Psi p$ and $\chi_{c1}
p$, respectively. These conditions imply  two equations, in order,
\begin{align}  
\label{240715.3}
1=&|\gamma_1|^2\left|\frac{\partial G_1(s_P)}{\partial s}\right|^2 + 
|\gamma_2|^2\left|\frac{\partial G_2(s_P)}{\partial s}\right|^2~,\\
\label{240715.4}
\Gamma_P=&\frac{q_1(m_P^2)|\gamma_1|^2}{8\pi m_P^2}+|\gamma_2|^2 \rho_2~.
\end{align}
The second term on the right-hand side (rhs) of the last equation  is the 
partial decay width of the $P_c(4450)$ to  $\chi_{c1}p$ and, since the
threshold of this channel is almost on top of the mass of the resonance, one should 
calculate it taking into account its finite width. 
  For that we follow the
procedure of Ref.~\cite{ref.240715.1} and  introduce a Lorentzian mass
distribution around the mass of the resonance due its width.  
 In this way, the formula for the partial decay width is recast as in the last
term on the rhs of Eq.~\eqref{240715.4} with
\begin{align}
\label{240715.5}
\rho_2= &\frac{1}{16\pi^2}\int_{m_1+m_2}^\infty dW\frac{q(W^2)}{W^2}
\frac{\Gamma_P}{(m_P-W)^2+\Gamma_P^2/4}~.
\end{align}
In practical terms we restrict ourselves to the resonance region so that we
integrate in this equation up to $m_P+2\Gamma_P$ as 
in Ref.~\cite{ref.240715.1}. Otherwise, the tail of the integrand takes too
long to converge. We have  checked that  in this way once the resonance mass
is above the $\chi_{c1} p$ threshold, within the one-sigma interval according
to the error in the mass  provided by Ref.~\cite{ref.230715.1} (we add
quadratically the statistical and systematic errors), the standard formula for the decay width 
(used in the first term on the rhs of Eq.~\eqref{240715.4} for the partial decay width to $J/\Psi p$) 
and the one based on Eq.~\eqref{240715.5} provide consistent results. However,
by making use of this procedure we avoid having a zero 
decay width to $\chi_{c1} p$ once $m_P$ is  smaller than the $\chi_{c1} p$
threshold. Furthermore, this would be an unphysical result. 

Equations  (\ref{240715.3},\ref{240715.4}) are valid for any partial wave or combination of partial waves in which each 
state $J/\Psi p$ or $\chi_{c1} p$ could be involved. For the case of the partial decay widths in Eq.~\eqref{240715.4} higher powers 
of three-momentum are reabsorbed in the residues squared and this is why the resulting expression seems the typical one for an $S$ wave decay.
  On the other hand,  independently of the quantum numbers 
to characterize a given partial wave, e.g. the basis $\ell S J$ (orbital angular momentum, total spin and total total angular momentum, respectively), 
the threshold is always the same and fixed by the particle masses.  As a result the derivative of $G_i(s_P)$ in Eq.~\eqref{240715.3} and 
the phase space driven factors in Eq.~\eqref{240715.4} do not depend on the specific partial wave and then each $|\gamma_i|^2$ 
  represents indeed the sum of residues squared to the involved partial waves.

Regarding the pole position $s_P$ of the $P_c(4450)$, the fact that the
threshold for $\chi_{c1}p$ is so close  to its  mass  makes necessary the
distinction on the Riemann sheet in which $s_P$ lies. 
In the following, we discuss  our results distinguishing between whether the
pole is located  in the 2nd or 3rd Riemann sheet. In the former, $\Ima
q_1<0,~\Ima q_2>0$, while in the latter $\Ima q_1<0,~\Ima q_2<0$. These are the 
two Riemann sheets that connect continuously with the physical axis below and
above the $\chi_{c1}p$ threshold, respectively. 
Notice that  $dG_2(s_P)/ds$ in Eq.~\eqref{240715.3}  depends on the actual
Riemann sheet taken for the pole position $s_P$.
Nevertheless, our results are rather stable under the change of sheet because
the calculation of $|\gamma_2|^2$ from Eq.~\eqref{240715.3} 
depends mainly on $\big|\partial G_2(s_P)/\partial s\big|^{-1}$, which is
rather stable under the change of sheet because the threshold of $\chi_{c1} p$ 
is very close to the mass of the relatively narrow resonance $P_c(4450)$. 
 Due to the latter reason it is also necessary  to solve
Eqs.~(\ref{240715.3},\ref{240715.4}) taking into account the error bars
in the mass and width of $P_c(4450)$ from 
Ref.~\cite{ref.230715.1}. 

\begin{table}[ht]
\begin{center}
\begin{tabular}{|l|cccccc|}
\hline
RS  & $|\gamma_1|$~(GeV) & $|\gamma_2|$~(GeV) & $|\Gamma_1|$~(MeV) & $|\Gamma_2|$~(MeV) & $X^f_1$ & $X^f_2$ \\
\hline
2nd &  1.9         &   15.6      &    5.6       &    34.6      & 0.004  & 0.996   \\   
    &  $\pm$0.5    & $\pm$1.3    &  $\pm$3.0    &  $\pm$ 10.9  & $\pm$0.002  & $\pm$0.002   \\
\hline
3rd &  2.4         &   14.6      &    9.5       &    30.0      & 0.006  &  0.994 \\
    &  $\pm$0.6    &   $\pm$1.1  &  $\pm$ 4.5   &  $\pm$9.0    & $\pm$0.003  &  $\pm$0.003 \\
\hline
\end{tabular}
{\caption {\small 
 Average couplings from the solution of
    Eqs.~\eqref{240715.3} and \eqref{240715.4} are shown in  columns 2,~3.  
    We also give the average partial decay widths $\Gamma_i$ (columns 4,~5)
    and resulting compositeness coefficients $X_i^f$ (columns 6,~7). 
    Below each quantity we provide its corresponding error. 
    The Riemann sheet (RS) where $s_P$ lies is given in the first column.
\label{tab.240715.1}}}
\end{center}
\end{table}

We present in Table~\ref{tab.240715.1} the results of solving 
Eqs.~(\ref{240715.3},\ref{240715.4}), where in the first 
column we show the Riemann sheet and then the couplings, partial decays widths
and final compositeness coefficients are 
given. As argued above, the channel $\chi_{c1} p$ has a much stronger coupling
to $P_c(4450)$ than to the lighter one, otherwise it would be an  
extremely wide resonance. The heavier channel is also by far the largest
component in our composite assumption for the  $P_c(4450)$
resonance. Regarding the partial width, we observe that the $\chi_{c1}p$ has a 
larger partial decay width than  the $J/\Psi p$. This fact is a rather robust
prediction of our model, because even if we reduced the weight of the two considered 
channels in the compositeness relation to $0.8$ instead of 1 in the lhs of Eq.~\eqref{240715.3}, 
still the partial decay width to  $\chi_{c1}p$ would be twice the one to $J/\Psi p$. 
This is shown in the second and third rows of Table~\ref{tab.240715.2}. For
definiteness, we give all  the results in this table in the  2nd Riemann sheet
since they are rather stable if changed to the 3rd Riemann sheet, 
similarly to the results shown above in Table~\ref{tab.240715.1}.

This observation could be turned around. If  the partial decay width of the
$P_c(4450)$ to either of the two channels were measured, but 
still assuming that the total width is the sum of these two partial decay
widths, Eq.~\eqref{240715.4}, one could find out whether the compositeness relation 
for the $P_c(4450)$ is saturated by the two channels $J/\Psi p$ and
$\chi_{c1}p$. This is illustrated in the last two rows of 
Table~\ref{tab.240715.2} where now  the lhs of  Eqs.~\eqref{240715.3} is 0.4.
As we can see, only when the $P_c(4450)$ has other large contributions in its 
composition beyond the two-body states $J/\Psi p$ and $\chi_{c1} p$, the
partial decay width to the former channel is the largest.

\begin{table}[ht]
\begin{center}
\begin{tabular}{|l|cccccc|}
\hline
$x$  & $|\gamma_1|$~(GeV) & $|\gamma_2|$~(GeV) & $|\Gamma_1|$~(MeV) & $|\Gamma_2|$~(MeV) & $X^f_1$ & $X^f_2$ \\
\hline
0.8 &  2.7                &   13.8            &   12.0             &    27.0            & 0.008  &  0.792   \\   
    &  $\pm$0.5  &    $\pm$1.2      &    $\pm$4.8   &     $\pm$8.5  & $\pm$0.003  &  $\pm$0.003   \\
\hline
0.4 &  4.0                &    9.6            &   26.0             &    13.0            & 0.016  &  0.384   \\
    &  $\pm$0.6           &    $\pm$0.8       &  $\pm$8.7          &  $\pm$ 4.0  & $\pm$0.005  &  $\pm$0.005   \\
\hline
\end{tabular}
{\caption {\small  Variation of the composite coefficient. The lhs of  Eqs.~\eqref{240715.3} is equal to 
                               $x$ (first column) and the pole lies in the 2nd Riemann sheet.
                               For further notation, see Table~\ref{tab.240715.1}
\label{tab.240715.2}}}
\end{center}
\end{table}
\medskip

\section{Summary}
\label{sec:sum}

In this work, we have analyzed the composite nature of the $P_c(4450)$ resonance measured by LHCb,
in a two-channel framework. The first, lower mass channel, is $J/\Psi p$ and the heavier one is $\chi_{c1} p$,
with its threshold extremely close to the mass of the resonance  \cite{Guo:2015umn}. We employ the present 
data on the relatively small width and mass of the resonance to conclude that within our assumption the $P_c(4450)$ is almost 
entirely a $\chi_{c1}p$ resonance, coupling much more strongly to this channel than to $J/\Psi p$, so that 
the former has clearly the largest decay width too.  As first noted here,
this is very similar to the scalar meson $f_0(980)$, that sits very close to the $K\bar K$ threshold  
because of its strong coupling to this channel. However, the coupling to $\pi\pi$ is much smaller which explains 
 its suppressed width, 
 though this lighter channel has plenty of phase space available. We have shown that this two-channel 
 composite nature of the $P_c(4450)$  can be tested by measuring precisely the partial widths into these two channels.

\subsection*{Acknowledgments}
JAO would like to thank the HISKP for its kind hospitality during a visit where most of these 
results were  obtained. This work is supported in part
by DFG and NSFC through funds provided to the Sino-German CRC 110
``Symmetries and the Emergence of Structure in QCD'' (NSFC Grant
No. 11261130311),  the Chinese Academy of Sciences CAS
President's International Fellowship Initiative (PIFI) grant no.
2015VMA076, the MINECO (Spain) and ERDF (European Commission) 
grant FPA2013-44773-P and the Spanish Excellence Network on Hadronic 
Physics FIS2014-57026-REDT.



\begin{thebibliography}{99}
\bibitem{ref.230715.1}  R.~Aaij {\it et al.} [LHCb Collaboration],
  arXiv:1507.03414 [hep-ex].


\bibitem{Chen:2015loa}
  R.~Chen, X.~Liu, X.~Q.~Li and S.~L.~Zhu,
  arXiv:1507.03704 [hep-ph].

\bibitem{Chen:2015moa}
  H.~X.~Chen, W.~Chen, X.~Liu, T.~G.~Steele and S.~L.~Zhu,
  arXiv:1507.03717 [hep-ph].

\bibitem{Roca:2015dva}
  L.~Roca, J.~Nieves and E.~Oset,
  arXiv:1507.04249 [hep-ph].

\bibitem{Feijoo:2015cca}
  A.~Feijoo, V.~K.~Magas, A.~Ramos and E.~Oset,
  arXiv:1507.04640 [hep-ph].

\bibitem{Mironov:2015ica}
  A.~Mironov and A.~Morozov,
  arXiv:1507.04694 [hep-ph].

\bibitem{Guo:2015umn}
  F.-K.~Guo, U.-G.~Mei{\ss}ner, W.~Wang and Z.~Yang,
  arXiv:1507.04950 [hep-ph].

\bibitem{Maiani:2015vwa}
  L.~Maiani, A.~D.~Polosa and V.~Riquer,
  arXiv:1507.04980 [hep-ph].

\bibitem{He:2015cea}
  J.~He,
  arXiv:1507.05200 [hep-ph].

\bibitem{Liu:2015fea}
  X.~H.~Liu, Q.~Wang and Q.~Zhao,
  arXiv:1507.05359 [hep-ph].

\bibitem{Lebed:2015tna}
  R.~F.~Lebed,
  arXiv:1507.05867 [hep-ph].




\bibitem{ref.240715.1} J.~A.~Oller and E.~Oset,
  Nucl.\ Phys.\ A {\bf 620} (1997) 438; (E) {\it ibid.} {\bf 652} (1999) 407 
  [hep-ph/9702314].

\bibitem{ref.230715.3}J.A.~Oller and Z.-H.~Guo, forthcoming. 

\bibitem{Weinberg:1962hj}
  S.~Weinberg,
  Phys.\ Rev.\  {\bf 130} (1963) 776.

\bibitem{Baru:2003qq}
  V.~Baru, J.~Haidenbauer, C.~Hanhart, Y.~Kalashnikova and A.~E.~Kudryavtsev,
  Phys.\ Lett.\ B {\bf 586} (2004) 53
  [hep-ph/0308129].


\bibitem{ref.230715.4}T.~Hyodo, D.~Jido and A.~Hosaka,
  Phys.\ Rev.\ C {\bf 85} (2012) 015201
  [arXiv:1108.5524 [nucl-th]].

\bibitem{ref.230715.5} F.~Aceti and E.~Oset,
  Phys.\ Rev.\ D {\bf 86} (2012) 014012
  [arXiv:1202.4607 [hep-ph]].

\bibitem{ref.230715.6}T.~Sekihara, T.~Hyodo and D.~Jido,
  PTEP {\bf 2015} 6,  063D04
  [arXiv:1411.2308 [hep-ph]].

\bibitem{Agadjanov:2014ana}
  D.~Agadjanov, F.-K.~Guo, G.~Rios and A.~Rusetsky,
  JHEP {\bf 1501} (2015) 118
  [arXiv:1411.1859 [hep-lat]].

\bibitem{Oller:2000fj}
  J.~A.~Oller and U.-G.~Mei{\ss}ner,
  Phys.\ Lett.\ B {\bf 500} (2001) 263
  [hep-ph/0011146].

\bibitem{ref.230715.7} 
  J.~A.~Oller,
  Eur.\ Phys.\ J.\ A {\bf 28} (2006) 63
  [hep-ph/0603134];




\end{thebibliography}
\end{document}